\documentclass[%
 reprint,
superscriptaddress,
 amsmath,amssymb,
 aps,pra
]{revtex4-2}

\usepackage{graphicx}
\usepackage{dcolumn}
\usepackage{bm}

\usepackage{subfigure}
\usepackage{xcolor}
\usepackage{amsmath}
\usepackage{physics}
\usepackage{float}
\begin{document}

\preprint{APS/123-QED}
\title{Probing the rotational spin-Hall effect in higher order Gaussian beams}

 \author{Ram Nandan Kumar}
 \affiliation{Department of Physical Sciences, Indian Institute of Science Education and Research Kolkata, Mohanpur-741246, West Bengal, India}

\author{Yatish}%
\affiliation{Department of Physical Sciences, Indian Institute of Science Education and Research Kolkata, Mohanpur-741246, West Bengal, India}

\author{Subhasish Dutta Gupta}
\affiliation{Department of Physical Sciences, Indian Institute of Science Education and Research Kolkata, Mohanpur-741246, West Bengal, India}
\affiliation{School of Physics, Hyderabad Central University, India}
\affiliation{Tata Institute of Fundamental Research Hyderabad, India}

\author{Nirmalya Ghosh}
\email{nghosh@iiserkol.ac.in}
\affiliation{Department of Physical Sciences, Indian Institute of Science Education and Research Kolkata, Mohanpur-741246, West Bengal, India}

\author{Ayan Banerjee}
 \email{ayan@iiserkol.ac.in}
 \affiliation{Department of Physical Sciences, Indian Institute of Science Education and Research Kolkata, Mohanpur-741246, West Bengal, India}

\date{\today}
\date{\today}

\begin{abstract}

Spin-to-orbit conversion of light is a dynamical optical phenomenon in non-paraxial fields leading to various manifestations of the spin and orbital Hall effect. However, effects of spin-orbit interaction (SOI) have not been explored extensively for higher order Gaussian beams carrying no intrinsic orbital angular momentum. Indeed, the SOI effects on such structured beams can be directly visualized due to azimuthal rotation of their transverse intensity profiles - a phenomenon we call the rotational Hall effect. In this paper, we show that for an input circularly polarized (right/left) $HG_{10}$ mode, SOI leads to a significant azimuthal rotation of the transverse intensity distribution of both the orthogonal circularly polarized (left/right) component, and the transverse component of the longitudinal field intensity with respect to the input intensity profile.  We validate our theoretical and numerically simulated results experimentally by tightly focusing a circularly polarized $HG_{10}$ beam in an optical tweezers configuration, and  projecting out the opposite circular polarization component and the transverse component of the longitudinal field intensity at the output of the tweezers. We also measure the rotational shift as a function of the refractive index contrast in the path of the tightly focused light, and in general observe a proportional increase. The enhanced spin-orbit conversion in these cases may lead to interesting applications in inducing complex dynamics in optically trapped birefringent particles using higher order Gaussian beams with no intrinsic orbital angular momentum.

\end{abstract}


\maketitle


Spin and orbital angular momentum  (SAM and OAM, respectively) are nearly independent quantities when light beams propagate paraxially in vacuum or through  isotropic and homogeneous media \cite{Int_1,Int_2,Int_3}, and are independently conserved during both paraxial and non-paraxial propagation \cite{Int_4}. However, propagation through inhomogeneous or anisotropic media \cite{Int_non_parax}, scattering processes \cite{Int_5}, as well as tight focusing in isotropic inhomogeneous media lead to interactions between SAM and OAM, giving rise to tangible effects in the mesoscopic scale with applications in nanosensing \cite{Int_5}, or particle manipulation \cite{the_3}.  In the case of tight focusing in optical tweezers, even a fundamental Gaussian beam evolves non-paraxially due to the presence of a spin-orbit and orbit-oribit interaction term in the expression of total angular momentum \cite{Int_5,Int_6,Int_7,Int_8,Int_9}. Now, both the spin-orbit interaction term (SOI) and orbit-orbit interaction terms are enhanced as a result of the generation of a large longitudinal component of the electric field due to tight focusing, and further accentuated by inserting a refractive index contrast in the path of the light beam that increases the geometric phase gradient of the focused light. On a different note, another interesting manifestation of the interplay between spin and orbital angular momentum arises in the spin and orbital Hall effect, which is essentially the transverse spatial separation of opposite angular momentum components induced by the spin-orbit and orbit-orbit interaction, preserving the angular momentum conservation law. This angular momentum dependent separation may be represented in terms of spin-spin separation called spin Hall effect \cite{Int_7,Int_10,Int_11,Int_12,Int_13,Int_14,Int_15}, orbit-orbit separation called orbital Hall effect \cite{Int_16,Int_17,Int_18}, and spin-orbit separation called spin-orbit Hall effect \cite{Int_19} - where the transverse spatial separation of the respective opposite spin and/or orbital angular momentum components is observed. 

The spin and orbital Hall effect have been extensively studied in both fundamental Gaussian beams \cite{the_3}, and in orbital angular momentum carrying Laguerre-Gaussian (LG) beams \cite{the_9,Int_1}, with interesting effects in spin-orbit and orbit-orbit conversions in tightly focused LG beams leading to intriguing orbital motion of single optically trapped birefringent particles \cite{the_8}. However, effects of spin-orbit interaction in structured Gaussian beams carrying no intrinsic orbital angular momentum (typically called Hermite-Gaussian or HG beams) have largely been ignored in the literature.  A study on such beams, however, merits attention, since any HG mode may be written as a superposition of two LG modes having opposite topological charge. This may lead to very interesting effects of spin-orbit interaction, especially in the case of input spin-polarized HG beams. In addition, the breaking of radial symmetry in the intensity profile distribution of such beams may lead to direct visualization of SOI effects in the beam structure itself.

In this paper, we study this problem in detail, and demonstrate clear manifestations of azimuthal rotation of the transverse intensity profile of  a circularly polarized $HG_{10}$ beam after it is tightly focused using optical tweezers. We describe this as a rotational spin-Hall effect. Thus, we observe that for input right or left-circularly polarized light (RCP and LCP, respectively), tight focusing using a high numerical aperture (NA) objective lens in an optical tweezers setup, leads to the generation of the opposite spin component that is also coupled with a corresponding orbital angular momentum mode - with the final effect being the rotation of the transverse intensity profile of the output.  For the same reason, the transverse component of the longitudinal field intensity generated due to non-paraxial propagation of the light, also displays a rotation.  The magnitude of rotation for the respective components is determined by the diffraction integrals (or Debye-Wolf integrals) $ I_{11}, I_{12}$ and $I_{14}$ (for transverse field components), and $I_{10}$ and $I_{13}$ (for longitudinal field components) - which provide the extent of the spin to orbit conversion of angular momentum \cite{Int_5, the_4}. We also study the dependence of the rotation on the refractive index (RI) contrast in the path of the beam after it is tightly focused, and observe that the rotational spin Hall effect in general increases monotonically with increasing RI contrast. We verify this experimentally by projecting out the opposite spin polarized transverse intensity component from the input, and the longitudinal component in an optical tweezers configuration. 

 We employ the Debye-Wolf theory or angular spectrum method \cite{the_2, the_4} to determine the electric field at the output of the high NA objective lens for an input spin-polarized HG beam, also considering RI stratification of the media through which the beam travels after focusing \cite{the_1,the_2,the_3,the_4,the_5}. The expression of the output electric field from the input electric field may be written as: 
\onecolumngrid
\begin{equation}
\left[\begin{array}{c}
E_{x}^{o} \\
E_{y}^{o} \\
E_{z}^{o}
\end{array}\right]=\left[\begin{array}{ccc}
i I_{11} \cos \psi+i I_{14} \cos 3 \psi & -i I_{12} \sin \psi+i I_{14} \sin 3 \psi & 2 I_{10}-2 I_{13} \cos 2 \psi \\
-i I_{12} \sin \psi+i I_{14} \sin 3 \psi & i\left(I_{11}+2 I_{12}\right) \cos \psi-i I_{14} \cos 3 \psi & -2 I_{13} \sin 2 \psi \\
-2 I_{10}+2 I_{13} \cos 2 \psi & 2 I_{13} \sin 2 \psi & i\left(I_{11}-I_{12}\right) \cos \psi
\end{array}\right] \times\left[\begin{array}{c}
E_{x}^{i} \\
E_{y}^{i} \\
E_{z}^{i}
\end{array}\right]
\label{eq1}
\end{equation}
\twocolumngrid
Where $E_{o}$ and $E_{i}$ denote the output and input electric fields related through the 3x3 Jones matrix, respectively, and $I_{10}, I_{11}, I_{12},  I_{14} $ and $I_{13} $  are the Debye-Wolf integrals. Now, given that the Jones vectors for $x$ and $y$-polarized input light are $\mathrm{E}_{\mathrm{x}}^i=\left[\begin{array}{lll}
1 & 0 & 0
\end{array}\right]^{T} \text { and } \mathrm{E}_{\mathrm{y}}^i=\left[\begin{array}{lll}
0 & 1 & 0
\end{array}\right]^{T}$, \\ 
we have from Eq.~\ref{eq1}, the output electric field for $HG_{10}$ input $x$-polarized light as
\begin{equation}
\left[\begin{array}{c}
E_{x}^{o} \\
E_{y}^{o} \\
E_{z}^{o}
\end{array}\right]_{x-pol}=\left[\begin{array}{c}
i I_{11} \cos \psi+i I_{14} \cos 3 \psi \\
-i I_{12} \sin \psi+i I_{14} \sin 3 \psi \\
-2 I_{10}+2 I_{13} \cos 2 \psi
\end{array}\right]
\label{eq2}
\end{equation}. It is important to note that the  circular polarized $HG_{10}$ beam does not carry OAM, but does possess an SAM of magnitude ±$\hslash$, so that the total angular momentum equals the SAM. Now, noting that the Jones vectors for input RCP and LCP light are $E_{r c p / l c p}=\left[\begin{array}{lll}
1 & \pm i & 0
\end{array}\right]^{T}
$, we use Eq.~\ref{eq1}, to determine the output electric field in both cases as
\begin{equation}
{\left[E_{x}^{0}~E_{y}^{0}~E_{z}^{0}\right]_{R C P / L C P}^{T}=\left[\begin{array}{lll}
a & b & c
\end{array}\right]^{\top}}, 
\label{eq3}
\end{equation}
where, $a=i I_{11} \cos \psi\pm I_{12} \sin \psi + I_{14} (i \cos 3\psi \mp \sin 3 \psi)$, $b=-i I_{12} \sin \psi+i I_{14} \sin 3 \psi \mp\left(I_{11}+2 I_{12}\right) \cos \psi \pm I_{14} \cos 3 \psi$, and $c=-2 I_{0}+2 I_{13} \cos 2 \psi \pm 2 i I_{13} \sin 2 \psi$.

\begin{center}
\begin{figure}[H]
\includegraphics[width=0.47\textwidth]{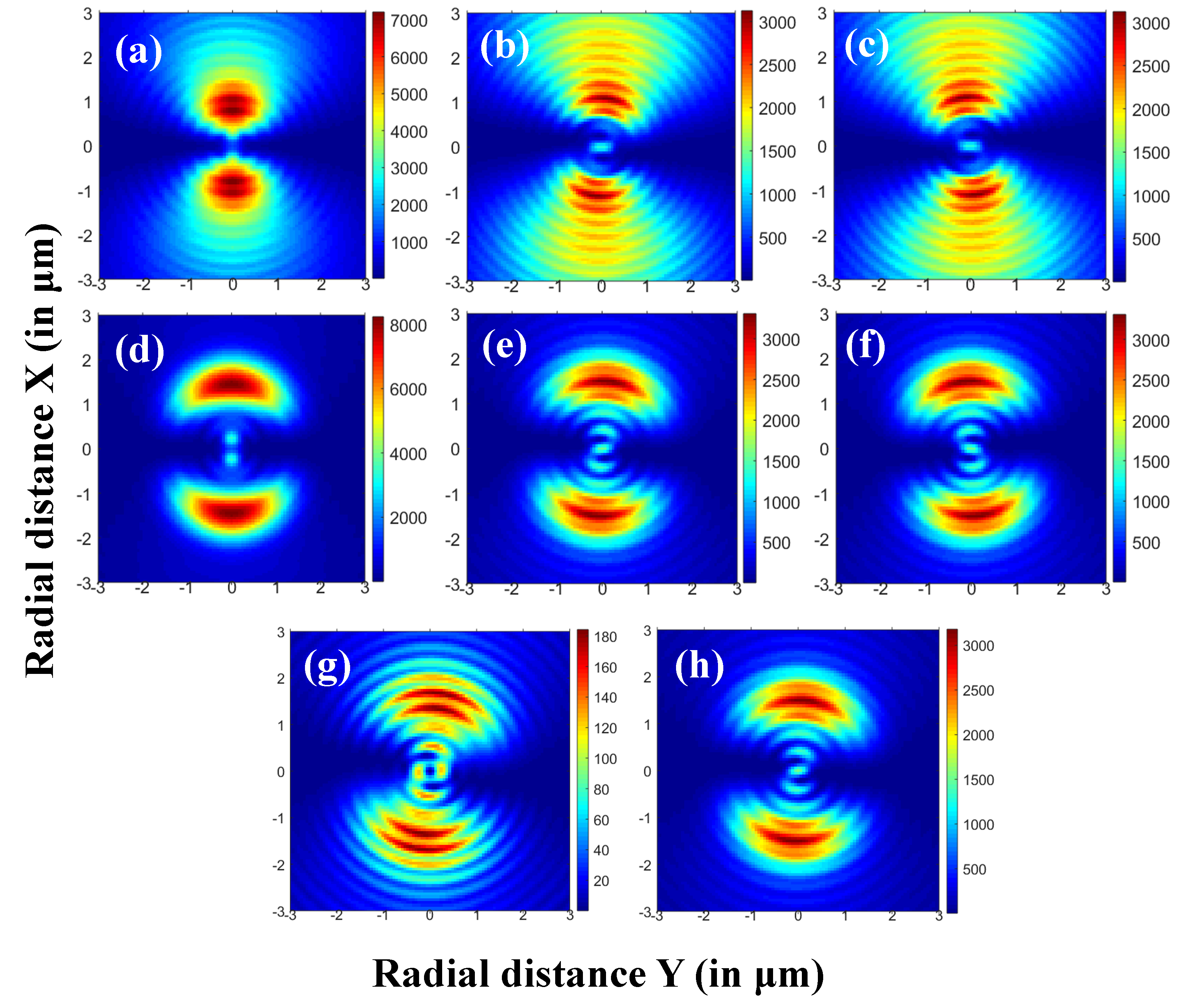}
\caption{Simulation of rotation of the total intensity profile at $z=2~\mu~m$ away the focus of the high NA objective (trap focus) for an input $HG_{10}$ beam mode for matched RI ((a), (b), (c)) and RI 1.814 ((d), (e) and  (f)). (a) and (d) are for input $x$-polarized light, while  (b) and (e) are for input RCP, (c) and (f) are for input LCP light, (g) rotation due to transverse components for RI 1.814, (h) rotation due to transverse component of longitudinal intensity profile for input RCP}
\label{ssphase0}
\end{figure}
\end{center}
We then decompose the above output electric fields for input RCP (LCP) light in terms of the SAM and OAM components - note that the effects of SOI cause all helicity components to be present along with the corresponding OAM modes in the output electric field.
\onecolumngrid
\begin{equation}
\left[\begin{array}{c}
E_{x}^{0} \\
E_{y}^{0} \\
E_{z}^{0}
\end{array}\right]_{R/L}=\frac{i}{2}\left(I_{11} e^{\pm i \psi}+I_{11} e^{\mp i \psi}+I_{12} e^{\pm i \psi}+I_{12} e^{\mp i \psi}\right)\\ \times \left[\begin{array}{c}
1 \\
\pm i \\
0
\end{array}\right] +i\left(I_{14} e^{\pm 3 i \psi}-I_{12} e^{\pm i \psi}\right)\left[\begin{array}{c}
1 \\
\mp i \\
0
\end{array}\right]\\+\left (2 I_{13}e^{\pm 2 i \psi }-2 I_{10}\right)\left[\begin{array}{c}
0 \\
0 \\
1
\end{array}\right]
\label{eq4}
\end{equation}
\twocolumngrid
It is clear that the first four terms are associated with the same helicity as the input light, coupled correspondingly with appropriate positive and negative OAM, respectively, while SOI leads to the fifth and sixth terms containing opposite helicity with $ l=\pm \hslash$ and $l=\pm 3\hslash $, respectively, satisfying the conservation of total angular momentum. The longitudinal component of the field with $l=\pm 2\hslash $ and $l=0$ appear in the last two terms. Understandably, the effects of SOI would be extracted by projecting out the field components having the opposite helicity as the input field, as well as the longitudinal component. This is manifested in the corresponding output intensity profile obtained from Eq.~\ref{eq4} - where, for an input RCP/LCP $HG_{10}$ mode, both the transverse (opposite spin to the input) and longitudinal intensity profiles display an azimuthal rotation about the input beam axis in a clockwise/anticlockwise direction (Fig.~\ref{ssphase0}). We observe these effects in our numerical simulations that we describe next.

We have simulated our experimental system, which consists of a stratified medium in the path of the tightly focused light after the output of the high NA objective in an optical tweezers configuration, and attempted to observe the rotational spin Hall-effect of light. As is well known, the linearly polarized light can be decomposed into RCP and LCP components in the transverse plane. Thus, the laser beam of wavelength 671 nm is incident on the 100× oil-immersion objective of NA 1.4 followed by (a) an oil layer of thickness around 5 $\mu m$ and refractive index (RI) 1.516, (b) a 160 $\mu m$ thick coverslip having refractive indices 1.516, 1.572, 1.695 and 1.814 (note that the case where the RI = 1.516 is henceforth referred to as the ``matched condition'', which is typically employed in optical tweezers to minimize spherical aberration effects in the focused beam spot.) (c) a water layer having a refractive index of 1.33 with a depth of 35 $\mu m$, and finally (chosen since probe particles in future experiments will be immersed in water) (d) a glass slide of refractive index 1.516 whose thickness we consider to be semi-infinite (~1500 $\mu m$). In the simulation, the origin of coordinates is taken inside the water layer at an axial distance of 5 $\mu m$ from the interface between water and the cover-slip. Thus, the objective-oil interface is at -170 $\mu m$, oil-coverslip interface is at -165 $\mu m$, cover-slip-sample chamber interface is at -5 $\mu m$, and sample chamber-glass slide interface is at +30 $\mu m$. On tight-focusing of the Hermite-Gaussian beam linearly polarized in the transverse plane, we observe that the electric field in the focal plane exhibits a component not only along the incident polarization direction, but also in the orthogonal and longitudinal directions. As we showed previously, the electric field in the transverse plane depends on Debye-Wolf integrals $I_{11}, I_{12},$ and $I_{14}$, while the longitudinal component depends on $I_{10}$ and $I_{13}$. 

\begin{center}
\begin{figure}[!h]
\includegraphics[width=0.35\textwidth]{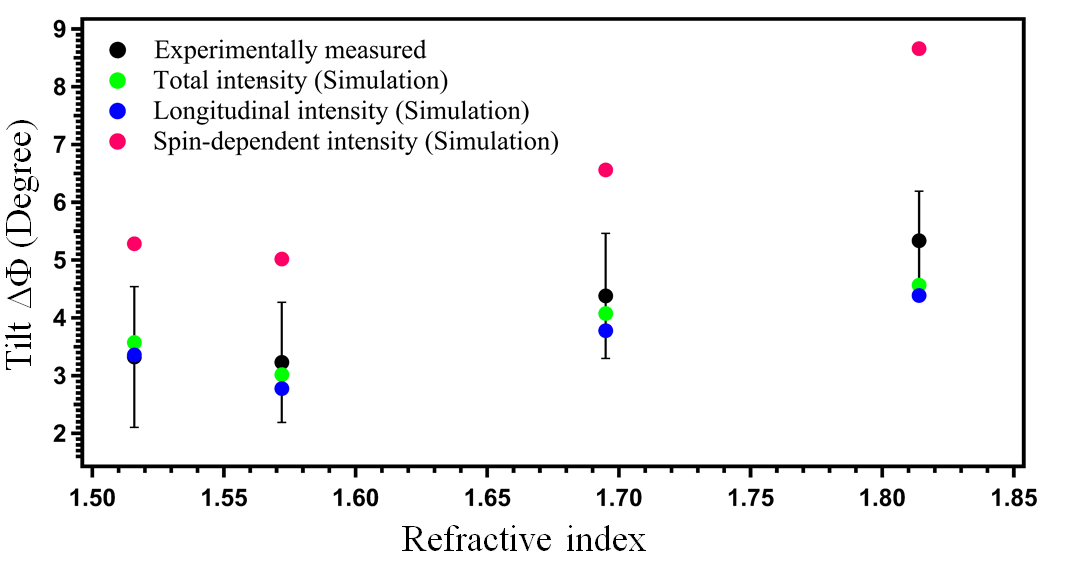}
\caption{ (a) Measurement of the rotational spin Hall effect for simulated transverse  (red solid circles), simulated longitudinal (blue solid circles), simulated total intensity (transverse + longitudinal, green solid circles)), and experimentally measured (transverse + longitudinal, black solid circles)) as a function of RI of the cover slip.}
\label{rotmeas}
\end{figure}
\end{center}

In order to extract the effects of SOI which lead to the rotational spin Hall effect, we plot the Debye-Wolf components associated with the polarization which is orthogonal to the input state, viz. RCP for input LCP, and vice versa. Thus, in Fig.~\ref{ssphase0}, we plot the transverse components ($I_{12}$ and $I_{14}$) of orthogonal polarization to the input state and longitudinal components of the total intensity (from Eq.~\ref{eq2}, \ref{eq4}) at $z=2~\mu$m away from the beam focus, and clearly observe rotations of the intensity profiles relative to the axis of the input modal distribution (rotational spin-Hall shift). We use both the matched RI for the cover slip (Fig.~\ref{ssphase0}(a)-(c)), and a mismatched RI of 1.814 (Fig.~\ref{ssphase0}(d)-(f)). While the rotational spin Hall shifts for input linear polarization ((Fig.~\ref{ssphase0}(a), (d)) is zero (since the shifts for the constituent RCP and LCP components cancel out), they are significantly high for both RCP ((Fig.~\ref{ssphase0}(b), (e)), and LCP (Fig.~\ref{ssphase0}(c), (f)). Note that the non-paraxial propagation of light also leads to a transverse component of the the longitudinal field intensity that also displays a rotation. We compare the rotational effects of the different field intensity components in Fig.~\ref{ssphase0}(g)-(h), where (g) shows the spin rotational shift due to the transverse components (LCP), and (h) shows the spin rotational shift due to the transverse component of the longitudinal intensity for input RCP light, and (e) the total (which term we now use to denote transverse LCP + transverse component of the longitudinal field intensities) shifts for input RCP light. We measure the rotations with respect to the axis of the linearly polarized beam mode that does not undergo rotation after focusing, and display the measured angles of rotation for an input RCP $HG_{10}$ beam as a function of RI from simulations in Fig.~\ref{rotmeas}. We observe a maximum of ~7.3 degrees azimuthal rotation for the transverse components for an RI of 1.814, while that for the transverse component of the longitudinal intensity profile is ~4.5 degrees. The total intensity - a combination of the rotations of both field intensity components - profile is observed to rotate by ~4.7 degrees azimuthally. We also observe that an increase in RI contrast tends to increase the observed azimuthal rotation. This can be understood from the fact that both the geometric phase acquired by the circularly polarized light during propagation, and its gradient, are enhanced with increasing RI contrast. Note that it is the geometric phase that is responsible for the SOI, while its gradient is at the heart of the observed spin-dependent azimuthal rotation of the intensity profile \cite{the_7}.  In addition, since the contribution of the transverse component of the longitudinal field intensity is ~90\% or even more of the total intensity (see Fig.~2 of Supplementary Information), it is the rotation of this component that finally dominates in the rotation of the total intensity profile, which we finally measure experimentally. We now proceed to describe our experiments to verify these simulations.

We use a conventional optical tweezers configuration consisting of an inverted microscope (Carl Zeiss Axioert.A1) with an oil-immersion 100× objective (Zeiss, NA 1.4) and a He-Ne laser (671 $nm$, 200 $mW$) coupled into the back port of the microscope. We use a vortex half-wave retarder of zero-order for generating a vector beam (i.e. LG beam). This beam is then passed through a linear polarizer to convert the LG mode into a $HG_{10}$ mode. Finally, we pass the beam through a quarter wave plate (QWP) centred at 671 $nm$ and oriented at 45 degrees to the input beam axis in order to circularly polarize the $HG_{10}$ mode. We then couple the circularly polarized HG beam into the microscope so that it is tightly focused into the stratified medium inserted beyond the objective lens. The stratified medium - as described earlier - consists of immersion oil (RI 1.516), cover slips (1.516, 1.572, 1.695 and 1.814 - we use individual cover slips for individual experiments), water (RI 1.33),  and a top glass slide (RI 1.516).

\begin{figure}
    \centering
    \includegraphics[width = 0.35\textwidth]{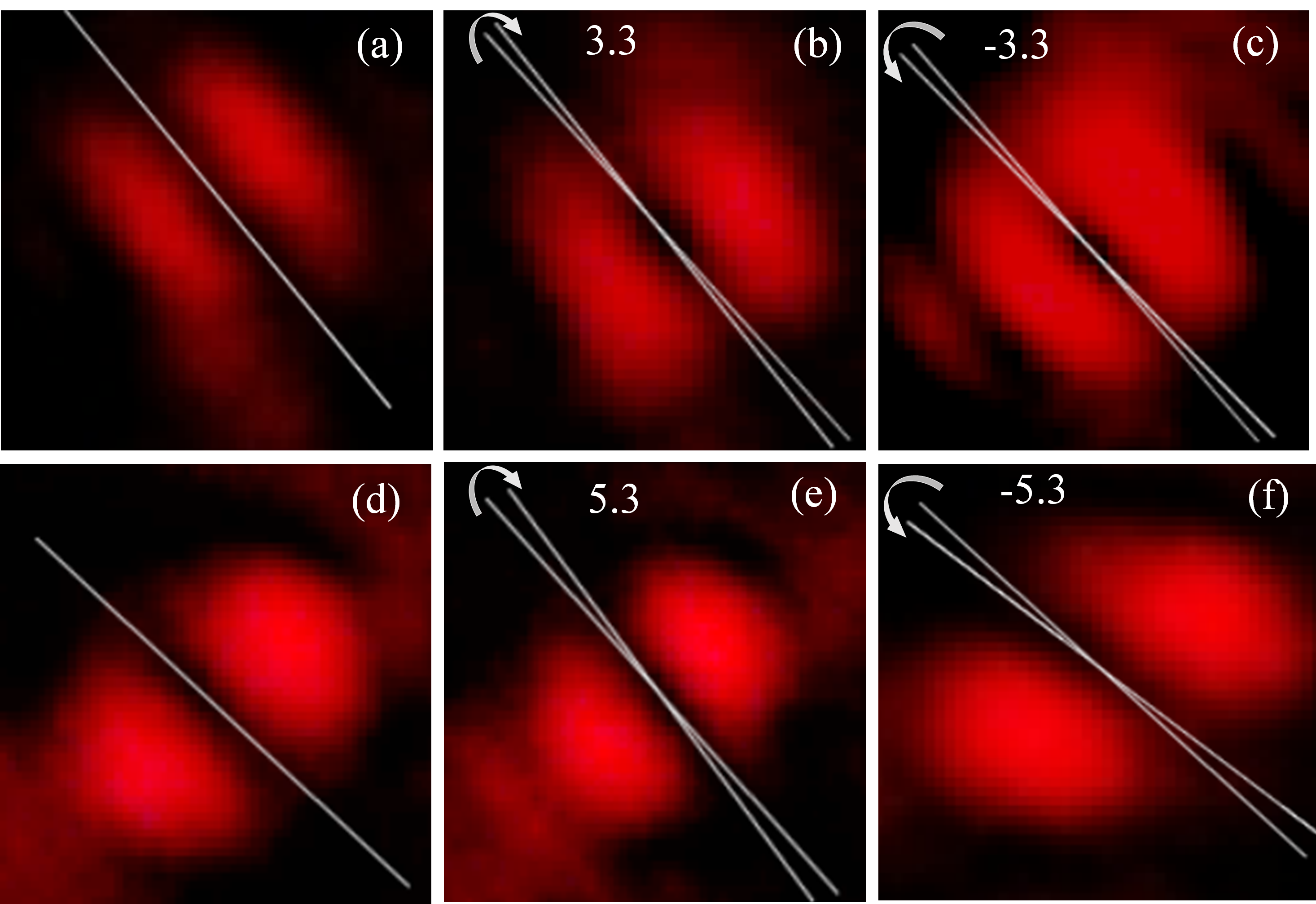}
    \caption{Experimental measurements of the rotational Hall shift for RI matched and mismatched cover slips. (a)-(c) are for matched conditions (RI 1.516) - with (a) CCD image of the output intensity profile for input linearly polarized state which we use as a reference to measure  rotations for input spin polarized states, (b) output intensity profile for input RCP, (c) LCP. (d)-(f) are for RI 1.814, with (d) reference linearly polarized ouput state, (e) output intensity profile for input RCP, (f) LCP. The white straight lines through the beam axes show the linear fit for lowest intensity pixels along the beam axis which we determine from a least squared fit in MATLAB.}
    \label{expresults}
\end{figure}

The back reflected light, from the cover slip passes through another QWP in order to project out the intensity of the spin state we would like to detect. Thus, for  incident RCP light, the output QWP filters out the RCP  component - so we detect the intensity only of the LCP component (superposed with the longitudinal component) on a CCD that we use to image the intensity profile. Importantly, tight focusing of linearly polarized HG beam does not show any rotation of the axis of $HG_{10}$ mode which we verify first. In order to measure rotation of the output mode, we perform the experiments first with linear polarized light that display no rotation, and can thus be employed as reference to measure the rotation for input spin polarized states. To determine the angle of rotation,  we use the CCD image of the mode to find out the distribution of pixels along the mode central region which have  minimum intensity, and fit a straight line connecting these pixels. This then defines the singularity axis of the mode (see Fig.~\ref{expresults}). We perform this analysis for all output modes for the input linear and spin polarized states, and measure the angle of rotation of the axis from the slope with respect to a vertical axis. We display the results of the rotation of the beam axes in Fig.~\ref{expresults}(a)-(f). Fig.~\ref{expresults}(a)-(c) display the rotational spin Hall shift for the matched RI, whereas (d)-(f) are for an RI of 1.814. Also, Fig.~\ref{expresults}(a) and (d) represent the reference linear polarization states with respect to which we measure the rotational spin Hall effect for the matched condition (Fig.~\ref{expresults} (b) and (c) - input RCP and LCP, respectively), and RI of 1.814 (Fig.~\ref{expresults} (e) and (f) - input RCP and LCP, respectively). Clearly, the beam axis appear tilted in opposite directions for opposite input spin states, with the tilt higher for the mismatched condition. The actual measured values of rotation are shown in Fig.~\ref{rotmeas} (black solid dots) - and we obtain a very good match with the values obtained from simulation (green solid dots). The errors in determining the rotation angles are between 5-20\% for individual cases. 

In conclusion, we study the rotational spin Hall effect in a structured Gaussian beam ($HG_{01})$ mode) carrying no intrinsic OAM, when it is tightly focused through a high NA objective lens in an optical tweezers configuration with a stratified medium in the path of the focused light. For input circularly polarized states, the SOI effects due to tight focusing cause the generation of opposite circular polarization coupled with a corresponding OAM state to satisfy conservation of total AM. As a result, the intensity profile corresponding to these states undergoes a rotation with respect to the axis of the input state, allowing us to visualize the effects of SOI in the output intensity profile itself. In addition, the transverse component of the longitudinal field intensity profile also undergoes a rotation due to coupling with OAM states due to SOI. We quantify such rotations using the complete vector diffraction theory described by Debye and Wolf by numerical simulations of the output electric field after tight focusing through a stratified medium, and proceed to verify our simulations via careful experiments, where we project out the opposite circular polarization state along with the longitudinal component for a given input circular polarization. We obtain excellent matches with simulation, and thus clearly demonstrate an interesting manifestation of SOI for higher order Gaussian modes of light. The rotational spin Hall shift increases with RI contrast of the stratified medium, which is expected since the magnitude of the SOI increases correspondingly due to the enhanced geometric phase gradient in these cases. We would also like to point out that, while these studies were carried out on the spatial intensity profiles of structured beams in an optical tweezers configuration, we would like to conduct future experiments on the observation of rotation of optically trapped birefringent particles around the beam axis due to the generation of OAM by SOI in such beams. These would thus lead to interesting routes of complex particle manipulation using optical tweezers, in which direction in our research is currently engaged. 

The authors acknowledge the SERB, Department of Science and Technology, Government of India (Project No. EMR/2017/001456) and IISER Kolkata IPh.D fellowship for research. They also acknowledge Anand Dev Ranjan and Sauvik Roy for their help in experiments and simulation.

\bibliographystyle{apsrev4-2}

\providecommand{\noopsort}[1]{}\providecommand{\singleletter}[1]{#1}%

\end{document}


\preprint{APS/123-QED}
\title{A study of the rotational spin-Hall effect in higher order Gaussian beams : SUPPLEMENTARY INFORMATION}

 \author{Ram Nandan Kumar}
 \affiliation{Department of Physical Sciences, Indian Institute of Science Education and Research Kolkata, Mohanpur-741246, West Bengal, India}

\author{Yatish}%
\affiliation{Department of Physical Sciences, Indian Institute of Science Education and Research Kolkata, Mohanpur-741246, West Bengal, India}

\author{Subhasish Dutta Gupta}
\affiliation{Department of Physical Sciences, Indian Institute of Science Education and Research Kolkata, Mohanpur-741246, West Bengal, India}
\affiliation{School of Physics, Hyderabad Central University, India}
\affiliation{Tata Institute of Fundamental Research Hyderabad, India}

\author{Nirmalya Ghosh}
\email{nghosh@iiserkol.ac.in}
\affiliation{Department of Physical Sciences, Indian Institute of Science Education and Research Kolkata, Mohanpur-741246, West Bengal, India}

\author{Ayan Banerjee}
 \email{ayan@iiserkol.ac.in}
 \affiliation{Department of Physical Sciences, Indian Institute of Science Education and Research Kolkata, Mohanpur-741246, West Bengal, India}

\date{\today}
\maketitle
\section{THEORETICAL CALCULATIONS}
The angular spectrum for the focal-field can be expressed as a function of the field at the focus as \cite{the_4}:
\begin{eqnarray}\label{polarint}
\vec{E}(\rho,\psi,z)&=&  i\frac{kfe^{-ikf}}{2\pi}\int_0^{\theta_{max}}\int_0^{2\pi}
\vec{E}_{res}(\theta,\phi)e^{ikz\cos\theta}\nonumber\\& \times & e^{ik\rho\sin\theta\cos(\phi-\psi)}
sin(\theta)\>d\theta d\phi,
\end{eqnarray}
where, $f$ is the focal length of the lens, and $\theta_{max}$ is because of the finite size of the aperture which is decided by the numerical aperture (NA) of the objective lens. Note that here we cannot use the paraxial approximation for calculating the electric field at the focal plane, because we are using a high NA objective lens (NA=1.4) for tight focusing. Thus, we use the angular spectrum method, which are the exact solutions for the non-paraxial regime to determine the nature of electric field distribution of Hermite Gaussian modes propagating in a medium or across stratified medium. This approach operates in the frequency domain, calculates the Fourier transform (FT) of the input field and multiplies the result with a transfer function, so that the desired output field is obtained by the inverse FT. The transfer function is given by $A=R_{z}(\phi) T R_{y}(\theta) R_{z}(-\phi)$ where $R_{z}$ and $R_{y}$ are $\mathrm{SO}(3)$ rotation matrices \cite{the_5}. Since the stratification of any medium makes the field propagating in the medium dependent on the input polarization, we incorporate the Fresnel transmission coefficients $\left(T_{s}\right.$
and $T_{p}$ ) as well as the Fresnel reflection coefficients $\left(R_{s}\right.$ and $\left.R_{p}\right)$ considering the contributions from both the $s$ and $p$ polarization. The output
and the incident field are related through a transfer function, A as:
$ \mathbf{E}_{r e s}(\theta, \phi)=A \mathbf{E}_{i n c}(\theta, \phi),$ where, the $T$ and $R$ matrices are given by:
\begin{equation}
T=\left(\begin{array}{ccc}
T_{p} & 0 & 0 \\
0 & T_{s} & 0 \\
0 & 0 & T_{p}
\end{array}\right) ; R=\left(\begin{array}{ccc}
-R_{p} & 0 & 0 \\
0 & R_{s} & 0 \\
0 & 0 & -R_{p}
\end{array}\right),
\end{equation}

Note that we have $ T_{i}^{(1, j)}=\dfrac{E_{i+}^{j}}{E_{i+}^{1}} ; R_{i}^{(1, j)}=\dfrac{E_{i-}^{j}}{E_{i+}^{1}};$. Here, $i$ specifies the polarization $(s$ and $p),~ +/-$ signifies a wave propagating forward and back-ward, respectively, and $j$ in the superscript specifies the layer of the stratified medium in which the optical tweezers (trapping laser) focus lies.

For an input $HG_{10}$ beam \cite{the_4}, we have
\begin{equation*}
\begin{array}{c}
E_{\text {inc }}(\theta, \phi)=E_{0}\left(2 x_{\infty} / w_{0}\right) \exp \left\{-\frac{x_{\infty}^{2}+y_{\infty}^{2}}{w_{0}^{2}}\right\}~~~~~~~~~~~~~~\\=\left(2 E_{0} f / w_{0}\right) \sin \theta \cos \phi e^{-f^{2} \sin ^{2} \theta / w_{0}^{2}}
\end{array}
\end{equation*}
From these equations, we can write the output electric field from the input electric field in the form of a matrix equation as:
\vspace*{1 cm}
\onecolumngrid
\begin{equation}
\left[\begin{array}{c}
E_{x}^{o} \\
E_{y}^{o} \\
E_{z}^{o}
\end{array}\right]=\left[\begin{array}{ccc}
i I_{11} \cos \psi+i I_{14} \cos 3 \psi & -i I_{12} \sin \psi+i I_{14} \sin 3 \psi & 2 I_{10}-2 I_{13} \cos 2 \psi \\
-i I_{12} \sin \psi+i I_{14} \sin 3 \psi & i\left(I_{11}+2 I_{12}\right) \cos \psi-i I_{14} \cos 3 \psi & -2 I_{13} \sin 2 \psi \\
-2 I_{10}+2 I_{13} \cos 2 \psi & 2 I_{13} \sin 2 \psi & i\left(I_{11}-I_{12}\right) \cos \psi
\end{array}\right] \times\left[\begin{array}{c}
E_{x}^{i} \\
E_{y}^{i} \\
E_{z}^{i}
\end{array}\right],
\end{equation}
where $E_{o}$ and $E_{i}$ denote the output and input Jones polarization vectors, respectively, and $I_{10}, I_{11}, I_{12},  I_{14} $ and $I_{13} $  are the Debye–Wolf integrals for the transmitted and reflected waves, which are given as \cite{the_1, the_2}
$$
\begin{array}{c}
I_{10}^{t}(\rho)=\int_{0}^{\theta_{\max }} E_{i n c}(\theta) \sqrt{\cos \theta} \sin ^{2} \theta T_{p} \sin \theta_{j} J_{0}\left(k_{1} \rho \sin \theta\right) e^{i k_{j} z \cos \theta_{j}} d \theta ,\\\\
I_{11}^{t}(\rho)=\int_{0}^{\theta_{\max }} E_{i n c}(\theta) \sqrt{\cos \theta} \sin ^{2} \theta\left(T_{s}+3 T_{p} \cos \theta_{j}\right) J_{1}\left(k_{1} \rho \sin \theta\right) e^{i k_{j} z \cos \theta_{j}} d \theta,\\\\
I_{12}^{t}(\rho)=\int_{0}^{\theta_{\max }} E_{i n c}(\theta) \sqrt{\cos \theta} \sin ^{2} \theta\left(T_{s}-T_{p} \cos \theta_{j}\right) J_{1}\left(k_{1} \rho \sin \theta\right) e^{i k_{j} z \cos \theta_{j}} d \theta,\\\\
I_{13}^{t}(\rho)=\int_{0}^{\theta_{\max }} E_{i n c}(\theta) \sqrt{\cos \theta} \sin ^{2} \theta T_{p} \sin \theta_{j} J_{2}\left(k_{1} \rho \sin \theta\right) e^{i k_{j} z \cos \theta_{j}} d \theta,\\\\
I_{14}^{t}(\rho)=\int_{0}^{\theta_{\max }} E_{i n c}(\theta) \sqrt{\cos \theta} \sin ^{2} \theta\left(T_{s}-T_{p} \cos \theta_{j}\right) J_{3}\left(k_{1} \rho \sin \theta\right) e^{i k_{j} z \cos \theta_{j}} d \theta,\\\\
I_{11}^{r}(\rho)=\int_{0}^{\theta_{\max }} E_{\text {inc }}(\theta) \sqrt{\cos \theta} \sin ^{2} \theta\left(R_{s}-3 R_{p} \cos \theta_{j}\right) J_{1}\left(k_{1} \rho \sin \theta\right) e^{-i k_{j} z \cos \theta_{j}} d \theta, \\\\
I_{12}^{r}(\rho)=\int_{0}^{\theta_{\max }} E_{\text {inc }}(\theta) \sqrt{\cos \theta} \sin ^{2} \theta\left(R_{s}-R_{p} \cos \theta_{j}\right) J_{1}\left(k_{1} \rho \sin \theta\right) e^{-i k_{j} z \cos \theta_{j}} d \theta, \\\\
I_{13}^{r}(\rho)=-\int_{0}^{\theta_{\max }} E_{\text {inc }}(\theta) \sqrt{\cos \theta} \sin ^{2} \theta R_{p} \sin \theta_{j} J_{2}\left(k_{1} \rho \sin \theta\right) e^{-i k_{j} z \cos \theta_{j}} d \theta, \\\\
I_{14}^{r}(\rho)=\int_{0}^{\theta_{\max }} E_{\text {inc }}(\theta) \sqrt{\cos \theta} \sin ^{2} \theta\left(R_{s}+R_{p} \cos \theta_{j}\right) J_{3}\left(k_{1} \rho \sin \theta\right) e^{-i k_{j} z \cos \theta_{j}} d \theta,
\end{array}
$$
Here, $t$ and $r$ superscripts denote the transmitted and reflected components, respectively, and $J_{n}$ is the Bessel function of the first kind of order $n$.\\

The radial intensity distribution of the output electric field with  input RCP and LCP HG beam is given by 
\onecolumngrid
\begin{equation}
\begin{array}{c}
I_{\text {RCP/LCP }}(\rho)=2\left(\left|I_{12}\right|^{2}+\left|I_{14}\right|^{2}\right)+2\left(\left|I_{11}\right|^{2}+\left|I_{12}\right|^{2}\right) \cos ^{2} \psi+2\left(I_{11} I_{12}^{*}+I_{11}^{*} I_{12}\right) \cos ^{2} \psi \\
-2\left(I_{12} I_{14}^{*}+I_{12}^{*} I_{14}\right) \cos 2 \psi \pm 2 i\left(I_{12} I_{14}^{*}-I_{12}^{*} I_{14}\right) \sin 2 \psi
\end{array}
\label{radialeqn}
\end{equation}
The equations for the intensity profile are different for input RCP and LCP $HG_{10}$ beams. Therefore, the rotation of the intensity profile for input RCP and LCP beams are in clockwise and anti-clockwise directions, respectively.
\section{NUMERICAL SIMULATION}
\begin{center}
\begin{figure}[]
\includegraphics[width=0.42\textwidth]{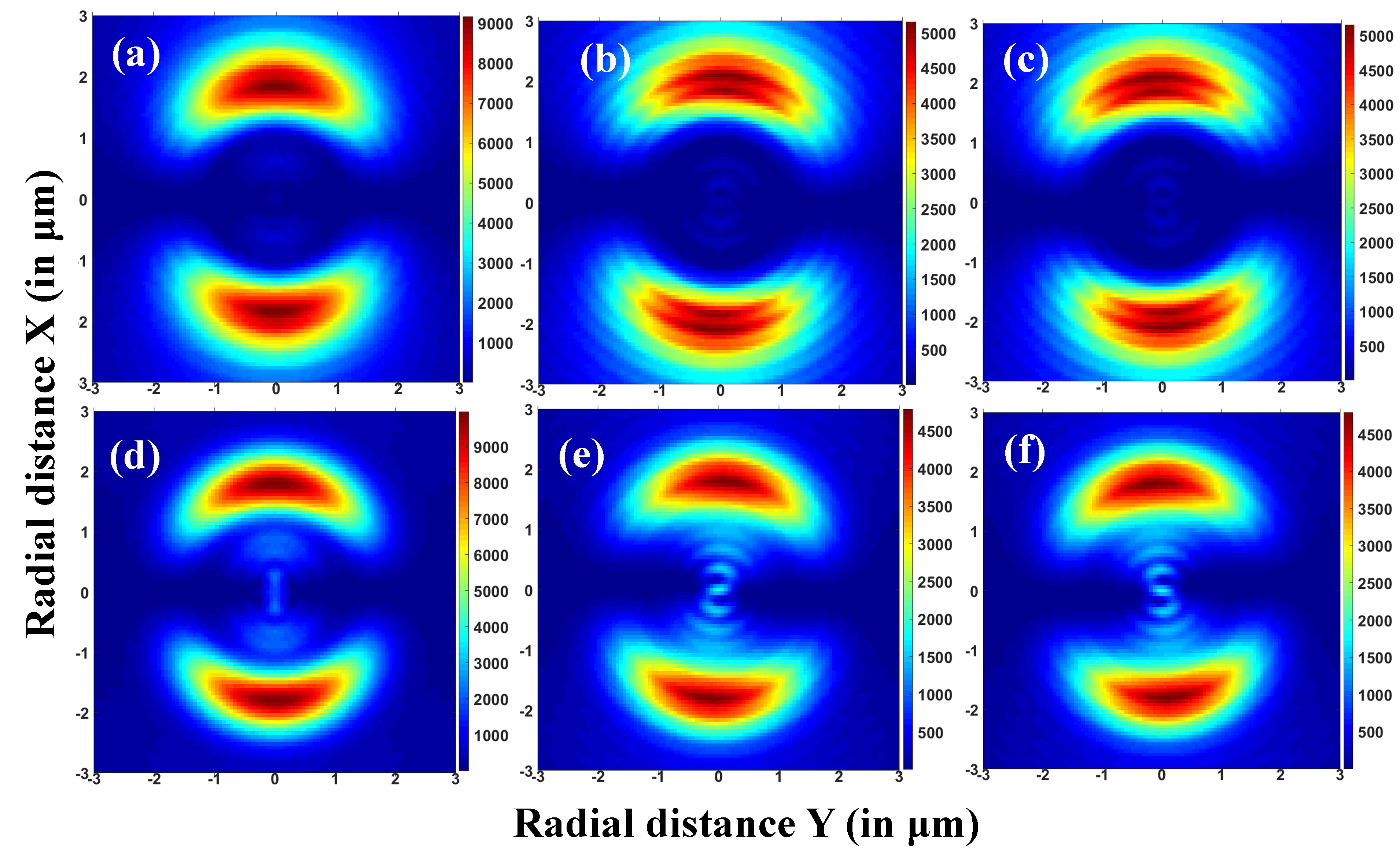}
\caption{Simulation of rotation of the total intensity profile at $z=2~\mu~m$ away the focus of the high NA objective (trap focus) for an input $HG_{10}$ beam mode for mismatched RI 1.572 ((a), (b), (c)) and RI 1.695 ((d), (e) and  (f)). (a) and (d) are for input $x$-polarized light, while  (b) and (e) are for input RCP, (c) and (f) are for input LCP light.}
\label{ssphase0}
\end{figure}
\end{center}

We now run simulations on our experimental system (stratified medium in the path of the optical tweezers light beam) as described in the primary manuscript. According to Eq.~\ref{radialeqn}, the transverse intensity of RCP and LCP components are different, therefore intensity profile for RCP and LCP input $HG_{10}$ mode show clockwise and anti-clockwise rotation with respect to x-axis in transverse plane. We had provided results for the azimuthal rotation for RI 1.516 and 1.814 (Fig.~1 of primary manuscript) - here we provide those for RI 1.572 and 1.659. Once again, we keep only the transverse field intensities corresponding to the circular polarization opposite to the input, and the transverse component of the longitudinal field intensity profile. The simulation results are displayed in Fig.~\ref{ssphase0}(a)-(f). The measurement of the rotation angles are performed using the method described in the primary manuscript.

\begin{center}
\begin{figure}[]
\includegraphics[width=0.42\textwidth]{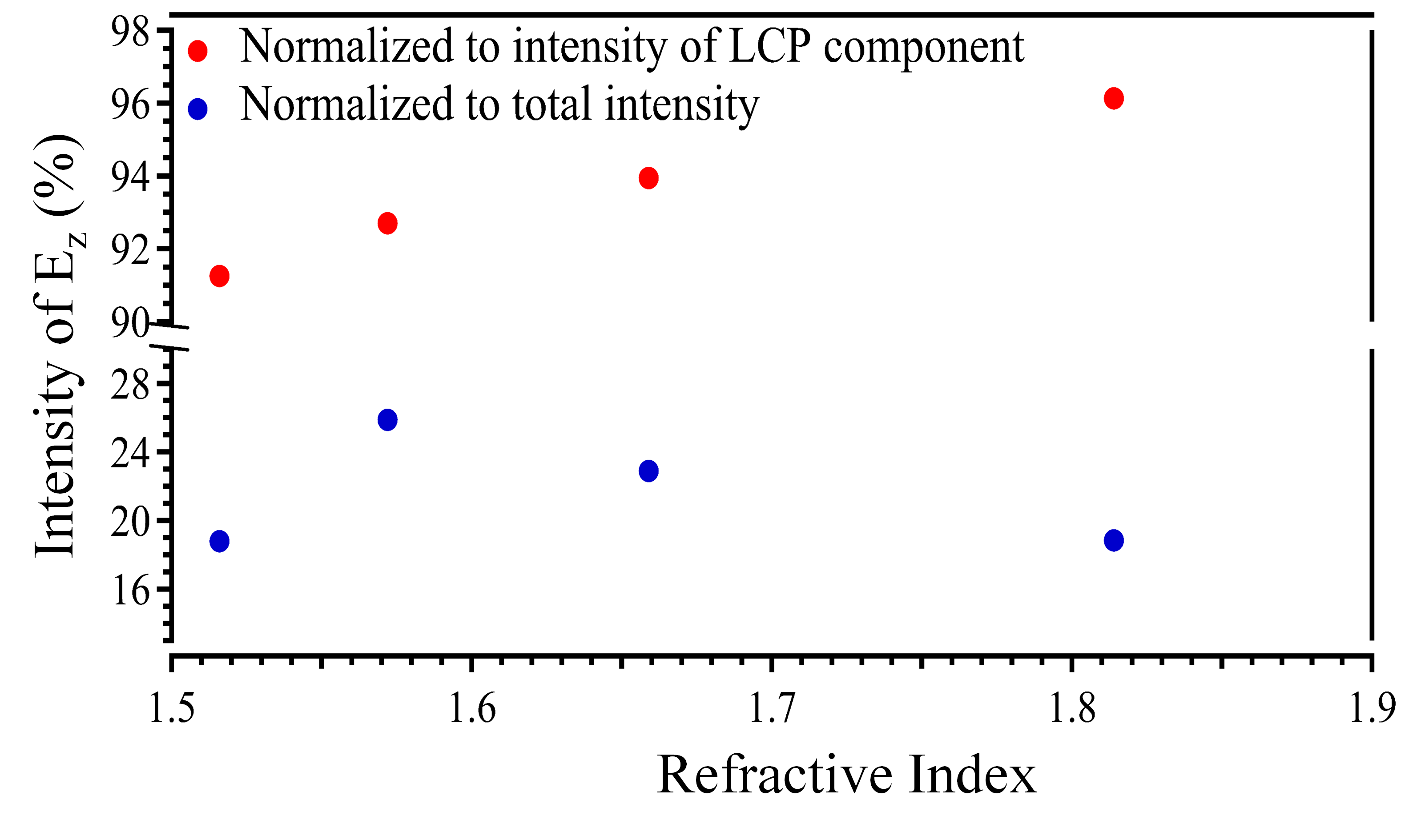}
\caption{ (b) Percentage of longitudinal component of light compared to  spin-polarized transverse component induced by SOI (red solid circles), and total transverse intensity (blue solid circles).}
\label{intensitycomp}
\end{figure}
\end{center}

Another useful exercise is to compare the transverse component of the longitudinal intensity profile ($E_{zt}$)to that of the transverse intensity component of the SOI-generated circular polarization that is opposite to the input polarization. This is displayed in Fig.~\ref{intensitycomp}. We observe that field corresponding to $E_{zt}$ (normalized) dominates over the transverse component of the radial intensity - with the former being almost 90\% of the sum of these two intensity components (refer to the total intensity defined in the text and in Fig.~1(g) of the primary manuscript) in the radial direction.  We also observe the contribution of the intensity profile of $E_{zt}$ to be increasing with increasing RI contrast of the stratified medium.  However, when we consider the total intensity in the radial direction, which also contains the intensity for the circular polarization in the direction of the input polarization, the $E_{zt}$ field intensity is only 20 to 30 \% of the total radial intensity. We also observe that this ratio reduces with increasing RI contrast, since the spread in the $k$ vector is reduced correspondingly, resulting in a lower z-component of the field compared to the total transverse component.
\begin{center}
\begin{figure}[!h!t]
\includegraphics[width=0.42\textwidth]{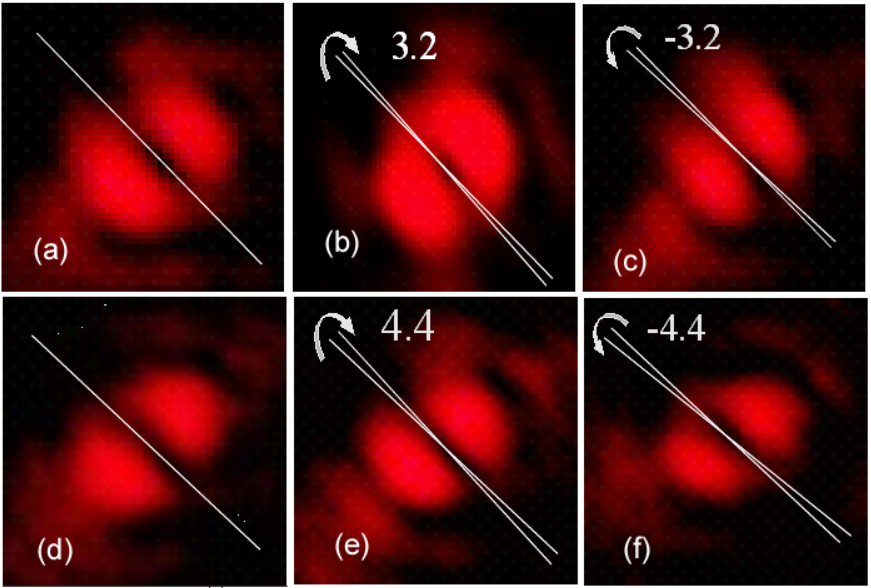}
\caption{Experimental measurements of the rotational spin Hall shift for RI mismatched cover slips. (a)-(c) are RI 1.572 - with (a) CCD image of the output intensity profile for input linearly polarized state which we use as a reference to measure rotations for input spin polarized states,(b) output intensity profile for input RCP, (c) LCP. (d)-(f) are for R.I 1.695 with (d) reference linearly polarized output state, (e) output intensity profile for input RCP, (f) LCP. The white straight lines through the beam axes show the linear fit for lowest intensity pixels along the beam axis.}
\label{exprotri1572}
\end{figure}
\end{center}
\section{Experiment}
We have also performed experiments to determine the azimuthal rotation for an input RCP $HG_{10}$ beam for RI 1.572 and 1.695 of the cover slip in the stratified medium. The beam profiles are displayed in Fig.~\ref{exprotri1572}(a)-(c) for RI 1.572 and (d)-(f) for RI 1.695. The measured rotations agree well with the theoretically simulated values as shown in Fig. 2 of the primary manuscript.

\bibliographystyle{apsrev4-2}

\providecommand{\noopsort}[1]{}\providecommand{\singleletter}[1]{#1}%